\documentclass[ aip, jmp, amsmath,amssymb, reprint, ]{revtex4-1}
\usepackage{graphicx}
\usepackage{dcolumn}
\usepackage{bm}
\usepackage{amssymb}
\usepackage{amsmath}
\usepackage{amsfonts}

\usepackage{hyperref}
\usepackage{bbm}
\usepackage[usenames]{color}
\begin{document}

\preprint{AIP/123-QED}

\title{Subcritical Andronov-Hopf scenario for systems with a line of equilibria}

\author{Ivan A. Korneev}
\affiliation{Institute of Physics, Saratov State University, Astrakhanskaya str. 83, 410012 Saratov, Russia}
\author{Andrei V. Slepnev}
\affiliation{Institute of Physics, Saratov State University, Astrakhanskaya str. 83, 410012 Saratov, Russia}
\author{Tatiana E. Vadivasova}
\affiliation{Institute of Physics, Saratov State University, Astrakhanskaya str. 83, 410012 Saratov, Russia}
\author{Vladimir V. Semenov}
\email{semenov.v.v.ssu@gmail.com}
\affiliation{FEMTO-ST Institute/Optics Department, CNRS \& University Bourgogne Franche-Comt\'e, 15B avenue des Montboucons, Besan\c con Cedex, 25030, France}

\date{\today}

\begin{abstract}
Using  numerical simulation methods and analytical approach, we demonstrate hard self-oscillation excitation in systems with infinitely many equilibrium points forming a line of equilibria in the phase space. The studied bifurcation phenomena are equivalent to the excitation scenario via the subcritical Andronov-Hopf bifurcation observed in classical self-oscillators with isolated equilibrium points. The hysteresis and bistability accompanying the discussed processes are shown and explained. The research is carried out on an example of a nonlinear memristor-based self-oscillator model. First, a simpler model including Chua's memristor with a piecewise-smooth characteristic is explored. Then the memristor characteristic is changed to a function being smooth everywhere. Finally, the action of the memristor forgetting effect is taken into consideration.
\end{abstract}

\maketitle

\begin{quotation}
The simultaneous dependence of the oscillatory dynamics both on parameter values and initial conditions is typical for systems with a line of equilibria. For this reason bifurcations without parameters are a frequent occurrence in such systems. In particular, memristor-based systems with a line of equilibria can exhibit oscillation excitation accompanied by various manifestations of the Andronov-Hopf bifurcation. The supercritical Andronov-Hopf bifurcation is well-studied in this regard. Meanwhile, the hard oscillation excitation associated with the subcritical Andronov-Hopf bifurcation has not been considered yet in the context of the systems with a line of equilibria. In the present paper we demonstrate the hard self-oscillation excitation in a system with a line of equilibria exhibiting distinctive features of the subcritical Andronov-Hopf bifurcation. The obtained results allow to carry out the comparative analysis of the hard self-oscillation excitation in classical self-oscillators with isolated equilibrium points and systems with a line of equilibria.
\end{quotation}

\section{Introduction}
\label{intro}

Depending on the dynamical system intrinsic peculiarities, the Andronov-Hopf bifurcation (also called the Poincar\'e-Andronov-Hopf bifurcation or simply the Hopf bifurcation) has different manifestations. Besides the classical bifurcation observed in deterministic self-oscillators, one can distinguish the stochastic Andronov-Hopf bifurcation occurring in the presence of noise or induced by random perturbations \cite{lefever1986,zakharova2010}, the delay-controlled and delay-induced bifurcations in deterministic and stochastic systems \cite{geffert2014,semenov2015-coh-res}, the bifurcations resulted from the piecewise-smooth character of the self-oscillator nonlinearity \cite{zhusubaliyev2003,simpson2018}, and other types. The particular kind of the Andronov-Hopf bifurcation discussed in the present paper is associated with the existence of a line of equilibria and involves both system parameters and initial conditions as factors controlling the bifurcation.

Dynamical systems can have $m$-dimensional manifolds of equilibria which consist of non-isolated equilibrium points (line of equilibria \cite{fiedler2000-1,fiedler2000-2,fiedler2000-3,liebscher2015}, surface of equilibria \cite{jafari2016}, circle \cite{gotthans2015,gotthans2016} and square \cite{gotthans2016} equilibria, etc). Among such manifolds one distinguishes normally hyperbolic manifolds of equilibria characterized by $m$ purely imagine or zero eigenvalues, whereas all the other eigenvalues have non-zero real parts. In the simplest case these manifolds exist as a line of equilibria. Systems with a line of equilibria have been mathematically considered\cite{fiedler2000-1,fiedler2000-2,fiedler2000-3,liebscher2015,riaza2012,riaza2016,corinto2020}. It has been shown that their significant feature is the occurrence of so-called bifurcations without parameters, i.e., the bifurcations corresponding to fixed parameters when the condition of normal hyperbolicity is violated at some points of the manifold of equilibria. 

\begin{figure}[t]
\centering
\includegraphics[width=0.47\textwidth]{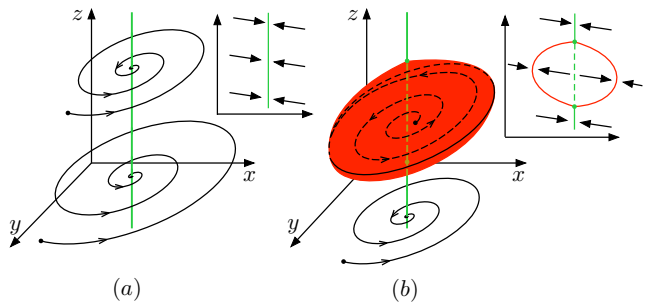} 
\caption{Illustration of self-oscillation excitation in three-dimensional systems with a line of equilibria associated with the supercritical Andronov-Hopf bifurcation. Before the bifurcation all trajectories are attracted to the line of equilibria (panel (a)). After the bifurcation all the trajectories are attracted either to stable semi-axes on the line of  equilibria or to invariant closed curves on a two dimensional surface (panel (b)). On all panels: the stable segments of the line of equilibria are marked by the solid green line while the dashed green line marks the unstable one. The attracting two-dimensional surface formed by the invariant closed curves is coloured in red. The right insets schematically show a section of the attractor by any plane being parallel to the OZ-axis and containing the line of equilibria (attraction is schematically shown).}
\label{fig1}
\end{figure}

Memristor-based oscillators are widely represented in a variety of systems with a line of equilibria. Bifurcation mechanisms of the periodic solution appearance in such systems have been explored numerically and analytically for different kinds of nonlinearity \cite{messias2010,botta2011,riaza2012,semenov2015,korneev2017,korneev2017-2}. In terms of the phase space, the oscillation excitation in the studied systems is associated with the invariant closed curve emergence in the neighbourhood of become unstable equilibrium points belonging to some part of the line of equilibria [Fig.~\ref{fig1}]. After that both a set of equilibrium points on the line of equilibria and a set of non-isolated closed curves can be observed depending on the initial conditions. All these trajectories create the system attractor which consists of some two-dimensional surface (the red surface in Fig.~\ref{fig1}~(b)) and stable branches of the line of equilibria (the green solid lines  in Fig.~\ref{fig1}~(b)).

Our previous studies of memristor-based systems with a line of equilibria have shown that some of them exhibit oscillation excitation with distinctive features of the supercritical Andronov-Hopf bifurcation \cite{korneev2017,korneev2017-2}. Thus, parallels between the bifurcation transitions in considered oscillators with a line of equilibria and self-oscillators with a finite number of isolated fixed points (for instance, the van der Pol self-oscillator) became visible. It has been revealed that the oscillations in systems with a line of equilibria corresponding to motions along invariant closed curves can be synchronized by external periodic forcing \cite{korneev2020} which increases the similarity. It allows to classify undamped periodic oscillations in autonomous systems with a line of equilibria as a special kind of self-sustained oscillations due to the possibility to observe the effect of frequency-phase locking.

Since the previously obtained results relate to the supercritical Andronov-Hopf bifurcation, the subcritical bifurcation became the next subject under study. In particular, a further series of questions arose after the demonstration of the similarity between self-oscillators with a finite number of steady states and systems with a line of equilibria. Is it possible to observe hard self-oscillation excitation in systems with a line of equilibria via transitions being equivalent to the subcritical Andronov-Hopf bifurcation? If yes then how the bistability (the coexistence of a stable steady state and a stable limit cycle after the saddle-node bifurcation) is realized in systems with a line of equilibria? Answers to all these questions are given in the current paper. 

\section{Models and methods}
\label{model}

A two-terminal passive electronic circuit element 'memristor' was introduced by Leon Chua \cite{chua1971} as a realization of the relationship between the electrical charge and the magnetic flux linkage. Then the term "memristor" was extended to the conception of "memristive systems" \cite{chua1976}.  A class of the memristive systems contains a broad variety of elements of different nature and is identified by a continuous functional dependence of characteristics at any time on previous states. 

Specifically, a flux controlled memristor relates the transferred electrical charge, $q(t)$, and the magnetic flux linkage, $\varphi(t)$: $dq=Wd\varphi$, whence it follows that $W=W(\varphi)=\dfrac{dq}{d\varphi}$. Using the relationships $d\varphi=V_{\text{m}}dt$ and $dq=I_{\text{m}}dt$ ($V_{\text{m}}$ is the voltage across the memristor, $I_{\text{m}}$ is the current passing through the memristor), the memristor current-voltage characteristic can be derived: $I_{\text{m}}=W(\varphi)V_{\text{m}}$. It means that $W$ is the flux-controlled conductance (memductance) and depends on the entire past history of $V_{\text{m}}(t)$:
 \begin{equation}
W(\varphi)=\dfrac{dq}{d\varphi}=q '\left( \int\limits_{-\infty}^{t}{V_{\text{m}}(t)dt} \right).
\label{W(phi)}
\end{equation}
The initially proposed by Leon Chua flux-controlled memristor \cite{chua1971} is described by piecewise-linear dependence $q(\varphi)$:
\begin{equation}
q(\varphi)=
\begin{cases}
	  (a-b)\varphi_{*}+b\varphi, & \varphi \ge \varphi_{*},\\
          a \varphi , & |\varphi| < \varphi_{*},\\
          -(a-b)\varphi_{*}+b\varphi , & \varphi \le -\varphi_{*}.
          
\end{cases}
\label{q_phi_chua_memristor}
\end{equation}
\begin{figure*}[t]
\centering
\includegraphics[width=0.84\textwidth]{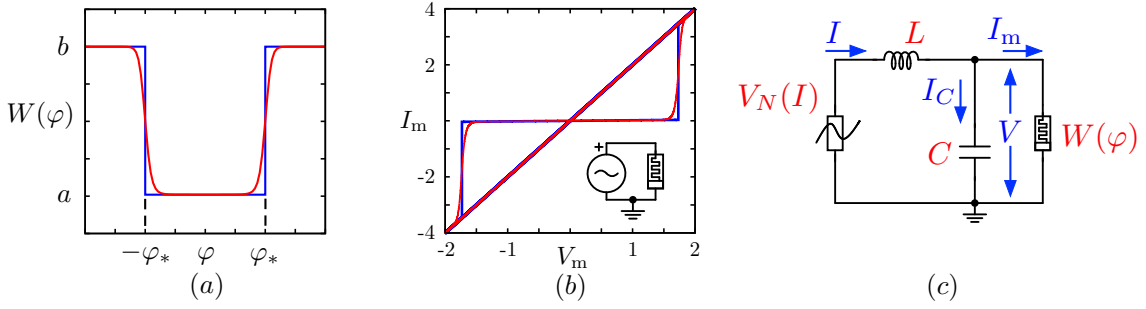} 
\caption{(a) Dependences of the memristor conductance on the state variable $\varphi$ according to formulas (\ref{chua_memristor}) (blue curve) and (\ref{tanh_memristor}) (red curve); (b) Current-voltage characteristic of the memristor driven by the periodic voltage signal $V_{\text{ext}}=V_{0}\sin(\omega_{\text{ext}}t)$ corresponding to memristor models (\ref{chua_memristor}) (blue loop) and (\ref{tanh_memristor}) (red loop). The memrisror parameters are $a=0.02$, $b=2$, $k=5$, $\varphi_{*}=1$. The applied voltage characteristics are $V_{0}=2$, $\omega_{\text{ext}}=1$; (c) Schematic circuit diagram of the model is under study (Eqs.(\ref{physical_system})).}
\label{fig2}
\end{figure*}  
Then the memristor conductance $W(\varphi)$ becomes:
\begin{equation}
W(\varphi)=
\begin{cases}
          a , & |\varphi| < \varphi_{*},\\
          b , & |\varphi| \geq \varphi_{*},
\end{cases}
\label{chua_memristor}
\end{equation}
which can be approximated by the hyperbolic tangent function (see Fig.~\ref{fig2}~(a)):
\begin{equation}
W(\varphi)=\dfrac{b-a}{2} \tanh\left(k(\varphi^2-\varphi_*)\right)+\dfrac{b+a}{2},
\label{tanh_memristor}
\end{equation}
where a parameter $k$ characterizes the sharpness of the transitions between two memristor's states. Changing the memristor conductance function to the smooth one does not qualitatively modify the memristor properties. The classical loop in the current-voltage characteristic of the memristor driven by the external periodic influence, $V_{\text{ext}}=V_{0}\sin(\omega_{\text{ext}}t)$, persists. It is illustrated in Fig.~\ref{fig2}~(b) where two memristor current-voltage characteristics are depicted for the set of parameter values $a=0.02$, $b=2$, $k=5$, $\varphi_{*}=1$, $V_{0}=2$, $\omega_{\text{ext}}=1$ (memristor models (\ref{chua_memristor}) and (\ref{tanh_memristor}) and their parameters are considered in the dimensionless form). It is important to note that the memristor model including tanh-nonlinearity is not the only smooth memristor model. There are many other smooth models describing various memristor properties \cite{tetzlaff2014,linn2014,singh2019,ascoli2013-2,chang2011,chua2011,guseinov2021}.

Real memristive devices can “forget” the state history over time. Thus, it has been shown that the “forgetting” effect in memristors based on metal oxides is associated with the diffusion of charged particles in a certain region with a high concentration of dopants \cite{chang2011,chen2013,zhou2019}. However, this “forgetting” can happen very slowly. One of the simplest form of the memristor state equation which implies the forgetting effect is the following:
\begin{equation}
\dfrac{dz}{dt}=x-\delta z,
\end{equation}
where $z$ plays a role of a memristor state variable, $x$ is an input signal (for example, it could be a voltage applied across the memristor), a parameter $\delta$ characterizes the forgetting effect strength.

In the current paper we consider a model of the electronic circuit depicted in Fig.~\ref{fig2}~(c). It is the series RLC-circuit including a nonlinear resistive element and a flux-controlled memristor connected in parallel with a capacitor. The nonlinear element is described by a S-type current-voltage characteristic, which is approximated by the fifth-order polynomial: $V_{N}(I)=-n_{1}I-n_{3}I^{3}+I^{5}$. The presented in Fig.~\ref{fig2}~(c) system is described by the following dynamical variables: $V$ is the voltage across the capacitor $C$,  $I$ is the current through the inductor $L$ and $\varphi$ is the magnetic flux linkage controlling the memristor. Using Kirchhoff’s laws, one can derive differential equations for the considered system:
\begin{equation}
\label{physical_system}
\left\lbrace
\begin{array}{l}
I=C\dfrac{dV}{dt'}+W(\varphi)V, \\
L\dfrac{dI}{dt'}+V_{N}(I)+V=0,\\
\dfrac{d\varphi}{dt'}=V,\\
\end{array}
\right.
\end{equation}
where $W(\varphi)$ is the flux-controlled conductance of the memristor. In the dimensionless variables $x=V / v_{0}$, $y=I/ i_{0}$ and $z=\varphi/(L\varphi_{0})$ with $v_{0}= 1$~V, $i_{0}=1$~A, $\varphi_{0}=\text{1 sec} \times v_{0}$ and the dimensionless time $t=[(v_{0}/(i_{0}L)]t'$, Eqs.(\ref{physical_system}) can be rewritten as being:
\begin{equation}
\left\lbrace
\begin{array}{l}
\dfrac{dx}{dt}=\alpha (y-G_{M}(z)x),\\
\\
\dfrac{dy}{dt} = - x +\beta_{1}y+\beta_{3}y^{3}-y^{5}, \\
\\
\dfrac{dz}{dt}=x,
\end{array}
\right.
\label{system}
\end{equation}
where $\alpha=(L/C)(i_{0}/v_{0})^{2}$  is the dimensionless parameter, which numerically equals to $L/C$. This parameter is assumed to be equal to unity in the following. The function $G_{M}(z)$ is the dimensionless equivalent of the function $W_{M}(\varphi)$. Two options for $G_M(z)$ are under consideration:
\begin{equation}
G_M(z)=
\begin{cases}
          a , & |z| < 1,\\
          b , & |z| \geq 1,
\end{cases}
\label{chua_memristor_dimensionless}
\end{equation}
and
\begin{equation}
G_M(z)=\dfrac{b-a}{2} \tanh\left(k(z^2-1)\right)+\dfrac{b+a}{2},
\label{tanh_memristor_dimensionless}
\end{equation}
where $a=0.02$ and $b=2$, $k=5$.

System (\ref{system}) is explored both theoretically by using the quasiharmonic reduction and numerically by means of modelling methods. Numerical simulations are carried out by integration of Eqs. (\ref{system}) using the fourth-order Runge-Kutta method with the time step  $\Delta t = 0.0001$ from different initial conditions.

\section{System with Chua's memristor}
\label{piecewise_smooth}

\subsection{Numerical simulation}
\label{piecewise_smooth_num_sim}
Let us consider system (\ref{system}) with memristor function (\ref{chua_memristor_dimensionless}). Parameters of system (\ref{system})-(\ref{chua_memristor_dimensionless}) are $\alpha=1$, $\beta_3=0.5$, $a=0.02$, $b=2$, $\beta_{1}\ge0$. The system has an infinite number of equilibrium points with the coordinates ($0;0;z\in(-\infty;+\infty)$) forming a line of equilibria in the phase space. One of the eigenvalues $\lambda_{i}$ of the equilibria is always equal to zero while the others depend on the parameters and on the position on the OZ-axis ($z$-coordinate):
\begin{equation}
\begin{array}{l}
\lambda_{1}=0, \\
\lambda_{2,3}=\dfrac{\beta_1-G_{M}(z)}{2} \pm \sqrt{\dfrac{(G_{M}(z)+\beta_1)^2}{4}-1}. 
\end{array}
\label{eigenvalues}
\end{equation}
\begin{figure}[t]
\centering
\includegraphics[width=0.47\textwidth]{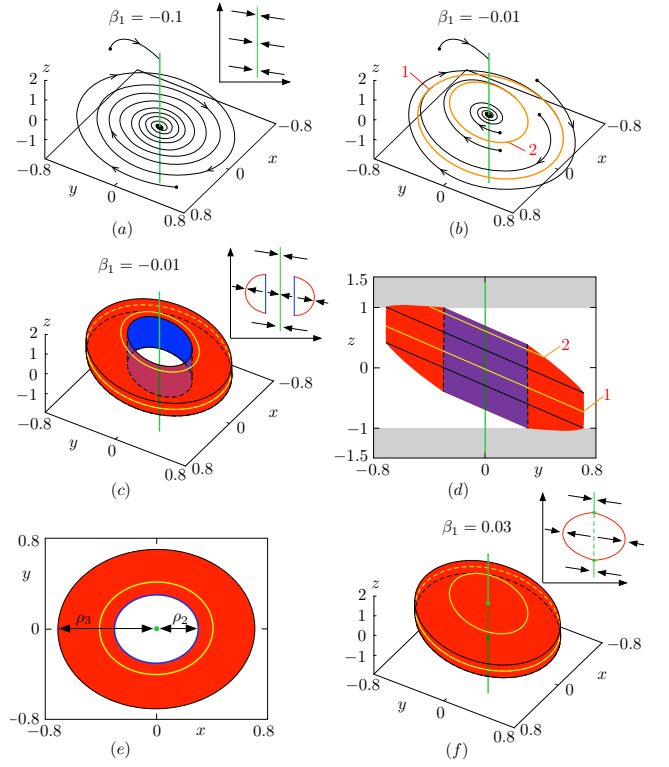} 
\caption{Evolution of system's (\ref{system})-(\ref{chua_memristor_dimensionless}) attractor caused by increasing parameter $\beta_1$. (a) $\beta_1=-0.1$: all trajectories tend to a line of equilibria being one single attractor; (b) $\beta_1=-0.01$: depending on the initial conditions the trajectories are attracted either to the line of equilibria or to a continuous manifold of closed curves (two of them are coloured in yellow); (c) Two attractors of the system at $\beta_1=-0.01$: the line of equilibria (coloured in green) and the manifold of invariant closed curves (coloured in red) separated by a two-dimensional cylindric surface (coloured in blue). Two closed curves from panel (b) are shown on the manifold (yellow curves); (d),(e) Projections of two attractors in panel (c) on the planes ($y$,$z$) and ($x$,$y$); (f) $\beta_1=0.03$: System has one attractor including two attracting semi-axes of the line of equilibria (steady states with the coordinates $x=y=0$,$|z|\ge1$) and a closed surface formed by a continuous manifold of  invariant closed curves. A manifold of unstable equilibria ($x=y=0$,$|z|<1$) is shown by a green dashed line. The right insets on panels (a), (c) and (f) schematically show a section of the attractors by any plane being parallel to the OZ-axis and containing the line of equilibria (attraction is schematically shown). Two yellow curves in panel (b) are marked on the attractive red manifold in panels (c)-(f).  Other system parameters are: $\beta_3=0.5$, $\alpha=1$, $a=0.02$, $b=2$.}
\label{fig3}
\end{figure}  
All the points belonging to the line of equilibria are neutrally stable along the OZ-axis. At the same time, for each point of the line of equilibria one can distinguish a plane $Q(x,y,z)$ which includes trajectories corresponding to attraction or repelling of this point in its vicinity. The stability of the equilibrium point belonging to the line of equilibria is chacterized by means of linear stability analysis used for isolated fixed points on the phase plane (analysis of the eigenvalues $\lambda_{2,3}$ in Exp. (\ref{eigenvalues})). Hereinafter, using the terms "stable" or "unstable" point in the line of equilibria, we mean the behavior of the trajectories in the neighbourhood of the equilibrium point on the plane $Q(x,y,z)$.

Increasing the parameter $\beta_1$ from the value $\beta_1=-0.1$ gives rise to the following bifurcation changes in the phase space. Initially, the system attractor is a line of equilibria, and all the phase trajectories are attracted to it (Fig. \ref{fig3}a). After passing through a certain value $\beta_1^{\text{SN}} \approx -0.036$ the second attractor appears in the phase space. Similarly to the previous configuration in Fig.~\ref{fig3}~(a), all the trajectories starting in a finite vicinity of the line of equilibria tend to it as before. The analysis of expressions (\ref{eigenvalues}) has confirmed the result that the line of equilibria is stable since all the equilibria ($x=y=0$, $z\in (-\infty; \infty)$) are characterized by two eigenvalues $\lambda_{2,3}$ with the negative real parts. At the same time, if one increases the distance between the initial condition point and the line of equilibria, the phase point can move away from the line of equilibria tracing a spiral-like trajectory. That movement culminates in motion along an invariant closed curve (two of them are coloured in yellow in Fig.~\ref{fig3}~(b)). Continuous changes of the initial conditions give rise to hit on another invariant closed curve. Thus, the second attractor represents a continuous set of closed curves which form a two-dimensional surface (the red surface in Fig.~\ref{fig3}~(c)) located in the subspace $z\in[-1;1]$ (see the corresponding projection in Fig.~\ref{fig3}~(d)). There exists the third limit set between two attractors. It is a two-dimensional cylindric surface (the blue surface in Fig.~\ref{fig3}~(c)) repelling the trajectories and separating the basins of attraction. Starting from the neighbourhood of the blue cylindric surface in Fig.~\ref{fig3}~(c) the phase point moves either to the line of equilibria or to the attracting manifold of closed curves (schematically shown on the right inset in Fig.~\ref{fig3}~(c)). The blue surface consists of a continuous set of invariant closed curves which can be observed by numerical modelling of system (\ref{system})-(\ref{chua_memristor_dimensionless}) with a negative time step $\Delta t<0$. 

Further growth of the parameter $\beta_1$ leads to the contraction of the repelling cylindric surface (radius $\rho_2$ in Fig.~\ref{fig3}~(e) decreases). Finally, the second bifurcation occurs at $\beta_1=0.02$ when the repelling surface collapses into a part of the line of equilibria ($x=y=0$, $|z|<1$). After that the equilibrium points with the coordinate $-1<z<1$ become unstable, while the points of the line of equilibria with  $|z|>1$ remain to be stable (the result of numerical simulation is confirmed by the analysis of eigenvalues (\ref{eigenvalues})). After the bifurcation the system has one single attractor again [Fig.~\ref{fig3}~(e)]. It consists of a continuous set of closed curves which form a two-dimensional surface (like a whirligig) for $-1<z<1$ and of stable semi-axes on the line of equilibria for $|z| \ge 1$.

\subsection{Analytical approach}
\label{piecewise_smooth_num_analytical}
To reveal the bifurcation mechanisms corresponding to the appearance and further transformation of attracting and repelling two-dimensional surfaces for $-1<z<1$, equations (\ref{system})-(\ref{chua_memristor_dimensionless}) are solved analytically. First, the system is transformed into the oscillatory form:

\begin{equation}
\label{chua_system_osc}
\left\lbrace
\begin{array}{l}
\dfrac{d^{2}x}{dt^{2}}+\left( G_{M}(z)-\beta_1 \right)\dfrac{dx}{dt}+\left( 1-\beta_1 G_{M}(z) \right)x\\
\\
=\beta_3\left( \dfrac{dx}{dt}+G_{M}(z)x\right)^{3} - \left( \dfrac{dx}{dt}+G_{M}(z)x\right)^{5}, \\
\\
\dfrac{dz}{dt}=x. \\
\end{array}
\right.
\end{equation}
The first equation of system (\ref{chua_system_osc}) does not include the variable $z$ in an explicit form. The dependence of the solution on the $z$ variable is defined by the function $G_{M}(z)$ which possesses two values. This allows to analytically solve the first equation (see system (\ref{chua_system_osc})) and then to consider two options for the function $G_{M}(z)$. The solution is sought as a harmonic function in complex representation: 
\begin{equation}
\label{x}
\left.
\begin{array}{l}
x(t)=Re \left\{ A(t)\exp{(i\omega_{0}t)}\right\}=\dfrac{1}{2}\{ A(t) \exp{(i \omega_{0}t)} \\
+A^{*}(t) \exp{(-i \omega_{0}t)} \} ,\\
\end{array}
\right.
\end{equation}
where $A(t)$ is the complex amplitude, $A^{*}(t)$ is the complex conjugate function, $\omega_{0}=\sqrt{1-\beta_1 G_{M}(z)}$ is the periodic solution frequency. To simplify further mathematical derivations, the complex amplitude and the corresponding complex conjugate function are written without arguments: $A(t) = A $, $A^{*}(t) = A^{*}$. The slowly varying amplitude condition is assumed to be satisfied: $\left| \dfrac{dA}{dt} \right| \ll \omega_{0}A$ and $\left| \dfrac{dA^{*}}{dt} \right| \ll \omega_{0}A^{*}$. Then the first and the second derivatives take the following form:
\begin{equation}
\label{dx}
\dfrac{dx}{dt}=\dfrac{1}{2}\left\{ i\omega_{0} A \exp {(i\omega_{0}t)} - i\omega_{0}A^{*} \exp {(-i\omega_{0}t)}    \right\} ,
\end{equation}
\begin{equation}
\label{ddx}
\left.
\begin{array}{l}
\dfrac{d^{2}x}{dt^{2}}= \left( i\omega_{0} \dfrac{dA}{dt} - \dfrac{\omega_{0}^{2}A}{2} \right) \exp{(i\omega_{0}t)}\\
\\
-\left( i\omega_{0} \dfrac{dA^{*}}{dt} + \dfrac{\omega_{0}^{2}A^{*}}{2} \right) \exp{(-i\omega_{0}t)}.\\
\end{array}
\right.
\end{equation}
Expressions (\ref{x})-(\ref{ddx}) are substituted into the first equation of system (\ref{chua_system_osc}). After that we approximate all fast oscillating terms by their averages over one period $T=2\pi/\omega_0$, which gives zero. Then we obtain the equation for the complex amplitude $A$:
\begin{equation}
\label{A}
\left.
\begin{array}{l}
\omega_0\dfrac{dA}{dt}=-\dfrac{i\left(\omega_0^2-1\right)+G_{M}(z)\omega_0}{2}A\\
\\
-\dfrac{\beta_1A}{2}\left( iG_{M}(z)-\omega_0 \right)\\
\\
-\dfrac{3\beta_3 A \left| A\right|^2}{8}\left(G_{M}^{2}(z)+\omega_0^2 \right)\left( iG_{M}(z)-\omega_0 \right)\\
\\
+\dfrac{5 A \left| A\right|^4}{16}\left(G_{M}^{2}(z)+\omega_0^2 \right)^2\left( iG_{M}(z)-\omega_0 \right).
\end{array}
\right.
\end{equation}
The representation of the complex amplitude in the form $A=\rho \exp(i \phi)$ (here $i$ is the imaginary unit, $i=\sqrt{-1}$) allows to rewrite Eq. (\ref{A}) as a system of equations for the real amplitude $\rho$ and phase $\phi$
\begin{equation}
\label{real_phase_and_amplitude}
\left.
\begin{array}{l}
\dfrac{d\rho}{dt}=\dfrac{\beta_1-G_{M}(z)}{2}\rho+\dfrac{3\beta_3}{8}\left( G_{M}^2(z)+\omega_0^2 \right)\rho^3\\
\\
-\dfrac{5}{16}\left( G_{M}^2(z)+\omega_0^2 \right)^2\rho^5,\\
\\
\dfrac{d\phi}{dt}=-\dfrac{\omega_0^2-1+G_M(z)\beta_1}{2\omega_0}\\
\\
-\dfrac{3\beta_3G_M(z)(G_M^2 (z)+\omega_0^2)}{8\omega_0}\rho^3\\
\\
+\dfrac{5G_M(z)(G_M^2 (z)+\omega_0^2)^2}{16\omega_0}\rho^5.
\end{array}
\right.
\end{equation}
It is important to note that the first equation does not include the phase $\phi$. Thus, the problem concerning the existence of periodic oscillations and bifurcations in system (\ref{chua_system_osc}) is reduced to the amplitude equation analysis.  The amplitude equation (the first equation in pair (\ref{real_phase_and_amplitude})) has three solutions:
\begin{equation}
\label{rho_123}
\left.
\begin{array}{l}
\rho_1=0,\\
\\
\rho_2=\sqrt{\dfrac{3\beta_3 - \sqrt{9\beta_3^2+40(\beta_1-G_M(z))}}{5\left( G_M^2(z) + \omega_0^2 \right)}},\\
\\
\rho_3=\sqrt{\dfrac{3\beta_3 + \sqrt{9\beta_3^2+40(\beta_1-G_M(z))}}{5\left( G_M^2(z) + \omega_0^2 \right)}}.
\end{array}
\right.
\end{equation}
The first equilibrium solution of the amplitude equation, $\rho_1=0$, exists at any values of parameters and corresponds to being at the equilibrium points $x=0, y=0, z\in(-\infty;+\infty)$. The solution $\rho_1=0$ is stable for $\beta_1<G_{M}(z)$ and unstable for $\beta_1>G_{M}(z)$. The second and third solutions (unstable $\rho_2$ and stable $\rho_3$) appear at $\beta_1=\beta_1^{\text{SN}}=G_M(z)-\dfrac{9}{40}\beta_3^2$. Increasing the parameter $\beta_1$ gives rise to decreasing the amplitude solution $\rho_2$ until it achieves the value $\rho_2=0$ at $\beta_1^{\text{HB}}=G_M(z)$.

\begin{figure}[t]
\centering
\includegraphics[width=0.25\textwidth]{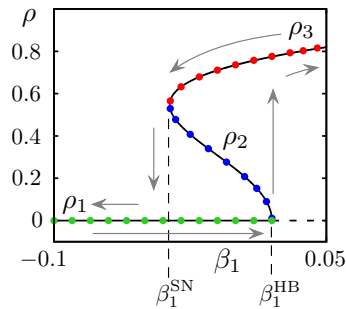} 
\caption{System (\ref{system})-(\ref{chua_memristor_dimensionless}) at $|z|<1$ where $G(z)=a$: analytically derived dependence of amplitudes, $\rho_{1,2,3}$ on the parameter $\beta_1$  (black solid lines, see Exps. (\ref{rho_123})) and the registered in numerical experiments
amplitude of oscillations $x(t)$. Two sets of the initial conditions were used for the numerical simulation: 
the neighbourhood of the origin ($x=y=z=0.01$) (the resulting oscillation amplitudes are coloured in green) and ($x=0.53$, $y=0.4$, $z=0$) (the resulting oscillation amplitudes are coloured in red). The initial condition ($x=0.53$, $y=0.4$, $z=0$) corresponds to motion along an invariant closed curve in the central segment of the attracting two-dimensional  surface (curve 1 in Fig.\ref{fig3} (b)-(f)). The amplitude of oscillations $x(t)$ corresponding to motions on the repelling two-dimensional surface (blue circles) was registered by the simulation of the system with the negative time step $\Delta t=-0.001$ starting from the point ($x=y=z=0.01$). System parameters are: $\beta_3=0.5$, $\alpha=1$, $a=0.02$, $b=2$.}
\label{fig4}
\end{figure}

Let us consider the evolution of amplitude equation solutions (\ref{rho_123}) in the subspace $-1<z<1$ at fixed $\beta_3=0.5$ and increasing $\beta_1$ [Fig.~\ref{fig4}]. In case $-1<z<1$ the function $G_M(z)$ takes the value $G_M(z)=a=0.02$. The appearance of the solutions $\rho_{2,3}$ at $\beta_1=\beta_1^{\text{SN}}=0.03625$ corresponds to the emergence of two non-isolated closed curves in the full phase space of system (\ref{system})-(\ref{chua_memristor_dimensionless}). A manifold of closed curves forming the attractive and repulsive two-dimensional surfaces appear simultaneously in the whole subspace $|z|<1$. The attractive two-dimensional surface consists of several parts. The central part is formed by a continuous manifold of identical invariant closed curves (one of them is the closed curve 1 in Fig. \ref{fig3}(b)-(f)). The central surface corresponds to the subpace, where the oscillation character is defined by the properties of the system fifth-order nonlinearity. In this case the instantaneous value $z(t)$ does not reach unity during the oscillatory process. The dependence of the oscillation amplitude $\rho_3$ on the parameter $\beta_{1}$ (see expressions (\ref{rho_123})) was derived for the central surface. If the value $|z(t)|=1$ is reached once a period of oscillations (see for example curve 2 in Fig. \ref{fig3}(b)-(f)), then the limitation of an oscillation amplitude is realized by the combined impact of the nonlinear element and the memristor. When the phase point moves out the subspace $-1<z<1$, the system dissipation sharply increases (for this regions $G_M(z) = 2$, see the grey areas in Fig.~\ref{fig3}~(d)). Then the amplitude growth stops. As a result, oscillations with constant amplitude are observed. 

Further growth of the parameter $\beta_1$ gives rise to decreasing the unstable solution amplitude $\rho_2$ up to $\rho_2=0$ while the stable solution amplitude $\rho_3$ continuously increases. This process corresponds to the contraction of the repelling cylindric surface (radius $\rho_2$ in Fig.~\ref{fig3}~(e) decreases). Finally, the second bifurcation occurs at $\beta_1^{\text{HB}}=G_M(z)=0.02$ when $\rho_2=0$ and the repelling surface collapses into the part of the line of equilibria ($x=y=0$, $|z|\le1$).

Arrows in Fig.~\ref{fig4} illustrate the hysteresis effect. Let us fix the parameter $\beta_1=-0.1$ and initial conditions close to the line of equilibria at $-1<z_0<1$. Then the phase point is attracted to the line of equilibria. Increasing the parameter $\beta_1$ does not shift the phase point from the line of equilibria until passing through the value $\beta_1=\beta_1^{\text{HB}}$. In case $\beta_1>\beta_1^{\text{HB}}$ the phase point moves away from the line of equilibria and comes to an invariant closed curve. If we keep the phase point on the invariant closed curve and decrease the parameter $\beta_1$, then the phase point continues to move along it until $\beta_1=\beta_1^{\text{SN}}$. When the parameter $\beta_1$ becomes smaller than $\beta_1^{\text{SN}}$, the phase point returns to the line of equilibria. The hysteresis principle is that the phase point moves from the line of equilibria and returns there at different values of $\beta_1$.

\begin{figure}[t]
\centering
\includegraphics[width=0.47\textwidth]{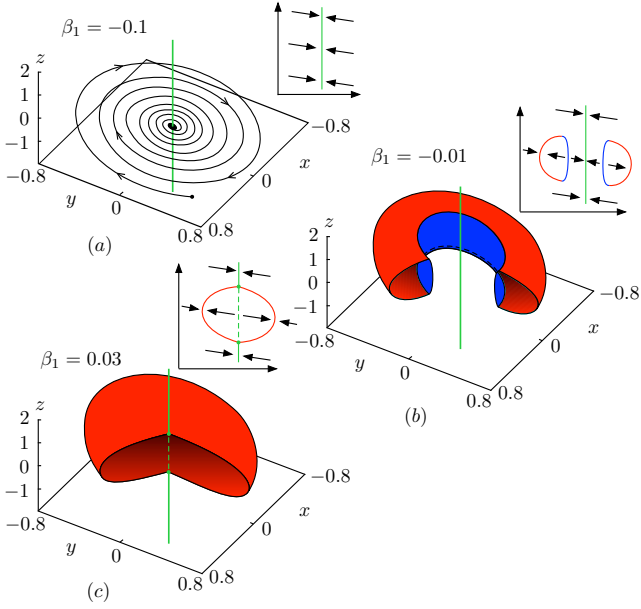} 
\caption{System (\ref{system}) including memristor characteristic function (\ref{tanh_memristor_dimensionless}): the attractor evolution caused by increasing parameter $\beta_1$ (to demonstrate surface sections, certain segments are cut off): $\beta_1=-0.1$ (panel (a)), $\beta_1=-0.01$ (panel (b)), $\beta_1=0.03$ (panel (c)). The colour scheme and designations are the same as in Fig. \ref{fig3}. System parameters are: $\beta_3=0.5$, $\alpha=1$, $a=0.02$, $b=2$, $k=5$.}
\label{fig5}
\end{figure}  

\section{System with smooth memristor nonlinearity}
\label{smooth}
The described above dynamics is associated with the piecewise-smooth memristor characteristic. The next stage is to understand how the attractor specifics changes in the presence of the smooth memristor nonlinearity. To explore this issue, system (\ref{system}) including memristor characteristic function (\ref{tanh_memristor_dimensionless}) is studied. 

It has been shown in numerical experiments that the system with smooth nonlinearity demonstrates the same bifurcation transitions (see Fig.~\ref{fig5}) in comparison with the piecewise-smooth oscillator. The difference consists in the shape of the attracting two-dimensional surface. Thus, one can conclude that the piecewise-smooth character of the nonlinearity does not play the principal role, and the described bifurcations can be observed in circuits including a wide range of memristors. The last issue is the impact of the memristor state equation which is studied in the further section. 

\section{Memristor forgetting effect}
\label{with_forgetting}

\begin{figure}[t]
\centering
\includegraphics[width=0.45\textwidth]{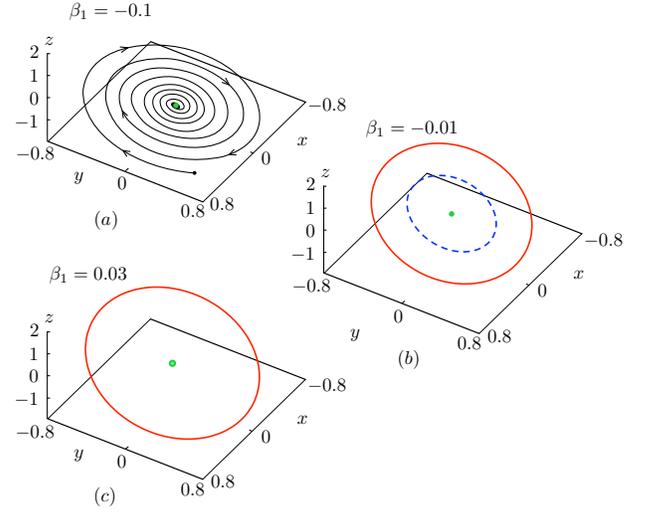} 
\caption{Evolution of system's (\ref{system_with_forgetting}) attractors caused by increasing parameter $\beta_1$: $\beta_1=-0.1$ (panel (a)), $\beta_1=-0.01$ (panel (b)), $\beta_1=0.03$ (panel (c)). Other system parameters are: $\beta_3=0.5$, $\alpha=1$, $a=0.02$, $b=2$, $k=5$, $\delta=0.01$. On all panels: the green steady state in the origin is a stable equilibrium in panels (a) and (b) and a saddle-focus in panel (c), the red solid curve is a stable limit cycle, the blue dashed curve is a saddle limit cycle.}
\label{fig6}
\end{figure}

Let us consider a model of the circuit in Fig.~\ref{fig2}~(c) including the memristor with the forgetting effect:
\begin{equation}
\left\lbrace
\begin{array}{l}
\dfrac{dx}{dt}=\alpha (y-G_{M}(z)x),\\
\\
\dfrac{dy}{dt} = - x +\beta_{1}y+\beta_{3}y^{3}-y^{5}, \\
\\
\dfrac{dz}{dt}=x-\delta z,\\
\\
G_M(z)=
\begin{cases}
          a , & |z| < 1,\\
          b , & |z| \geq 1,
\end{cases}
\end{array}
\right.
\label{system_with_forgetting}
\end{equation}
where parameters are: $\alpha=1$, $\beta_3=0.5$, $\delta=0.01$, $a=0.02$, $b=2$. In such a case the line of equilibria is transformed into a steady state $x=y=z=0$ being one single steady state in the phase space at $\beta_1< \beta^{\text{SN}}_1$ where $\beta^{\text{SN}}_1=a-\dfrac{9}{40}\beta_3^2$ [Fig.~\ref{fig6}~(a)]. Increasing the parameter $\beta_1$ gives rise to the saddle-node bifurcation of limit cycles at $\beta_1=\beta_1^{\text{SN}}$: a pair of limit cycles (stable and saddle ones) appear in the phase space [Fig.~\ref{fig6}~(b)]. Further growth of the parameter $\beta_1$ leads to the contraction of the saddle limit cycle which collides into the stable steady state at $\beta_1=a$. At this moment the steady state becomes a saddle-focus fixed point. After the bifurcation the stable limit cycle is a single attractor in the phase space  [Fig.~\ref{fig6}~(c)]. In this way, the attracting and repelling surfaces built by invariant closed curves in the phase space of systems with a line of equilibria are transformed into the stable and saddle limit cycles in the presence of the forgetting effect. Thus, one deals with the contraction of limit sets along the OZ-axis.

The action of the parameter $\delta$ is reflected in the duration of transient processes describing motions of the phase point to attractors along the OZ-axis. The smaller is the parameter $\delta$ the longer is the transient time.

\section*{Conclusions}
\label{conclusions}
The classical self-oscillation excitation scenario via the subcritical Andronov-Hopf bifurcation \cite{andronov1966} implies the saddle-node bifurcation of limit cycles which does not affect a stable steady state: stable and unstable limit cycles appear in the phase space. After that the unstable limit cycle shrinks and collides into the steady state which losses stability (the subcritical Andronov-Hopf bifurcation). After that the stable limit cycle becomes a single attractor in the phase space. It is shown in the current paper how such bifurcation transitions morph in systems with a line of equilibria. 
In such a case an image of the saddle-node bifurcation of limit cycles is the appearance of two two-dimensional surfaces formed by invariant closed curves. The first one becomes the second attractor in addition to the line of equilibria. The second surface is repelling and plays a role of an unstable limit cycle: it splits the basins of attraction of two attractors. Further change of parameter values leads to the contraction of the repelling surface until it collides into a part of the line of equilibria. At this moment the contacting segment of the line equilibria becomes unstable. After that two attractors merge into one which consists of two attracting semi-axes of the line of equilibria (steady states with the coordinates $x=y=0$,$|z|\ge1$) and a closed surface formed by a continuous manifold of invariant closed curves. 

To combine numerical simulations with analytical derivations, a simplified model with piecewise-smooth memristor nonlinearity was considered first. Further introduction of a continuous memristor characteristic has not caused principal changes in the structure of attractors and bifurcation transitions. 

The impact of the memristor forgetting effect is manifested in the contraction of limit sets along the OZ-axis: the line of equilibria morphs into a steady state while the repelling and attracting surfaces are transformed into unstable and stable limit cycles. Resultantly, a continuous dependence of the oscillation characteristics on the initial condition $z_0$ disappears. In such a case one deals with well-known hard self-oscillation excitation observed in self-oscillators with a finite number of isolated steady states and limit cycles. A similar character of the forgetting effect action was noted in earlier publications \cite{korneev2020-2,korneev2021} where the effect of synchronization involving a line of equilibria was studied.

A manifold of systems with a line of equilibria is not limited by memristor-based oscillators. It includes electronic circuits where the inertial nonlinearity of a memristive element is implemented by means of analog electronics (see, for instance, papers \cite{semenov2015,wang2017-2} where the memristor and memcapacitor characteristics are implemented). Generally, the reported bifurcation transitions are expected to be observed in systems with a line of equilibria of any nature.

\begin{acknowledgements}
The reported study was funded by the Russian Foundation for Basic Research (project No. 19-32-90030). 

We dedicate this paper to the memory of Vadim Semenovich Anishchenko who gave us advice in 2014 to focus on the dynamics of memristor-based circuits from the theoretical point of view. It has facilitated our initial steps in this field and stimulated the first publication appearance.
\end{acknowledgements}

\section*{DATA AVAILABILITY}
The data that support the findings of this study are available from the corresponding author upon reasonable request.


\end{document}